\documentstyle[]{mn}

\begin{document}

\def\spose#1{\hbox to 0pt{#1\hss}}
\def\simlt{\mathrel{\spose{\lower 3pt\hbox{$\mathchar"218$}}
     \raise 2.0pt\hbox{$\mathchar"13C$}}}
\def\simgt{\mathrel{\spose{\lower 3pt\hbox{$\mathchar"218$}}
     \raise 2.0pt\hbox{$\mathchar"13E$}}}
\def\eg{{\rm e.g. }}
\def\ie{{\rm i.e. }}
\def\etal{{\rm et~al. }}

\def\aj{{AJ}}			
\def\araa{{ARA\&A}}		
\def\apj{{ApJ}}			
\def\apjs{{ApJS}}		
\def\apss{{Ap\&SS}}		
\def\aap{{A\&A}}		
\def\aapr{{A\&A~Rev.}}		
\def\aaps{{A\&AS}}		
\def\azh{{AZh}}			
\def\jrasc{{JRASC}}		
\def\mnras{{MNRAS}}		
\def\pasa{{PASA}}		
\def\pasp{{PASP}}		
\def\pasj{{PASJ}}		
\def\sovast{{Soviet~Ast.}}	
\def\ssr{{Space~Sci.~Rev.}}	
\def\zap{{ZAp}}			
\def\nat{{Nature}}		
\def\aplett{{Astrophys.~Lett.}}
\def\fcp{{Fund.~Cosmic~Phys.}}
\def\memsai{{Mem.~Soc.~Astron.~Italiana}}
\def\nphysa{{Nucl.~Phys.~A}}
\def\physrep{{Phys.~Rep.}}

\title[Infall models, ICM enrichment \& the IMF]{Infall models of
elliptical galaxies: further evidence for a top-heavy initial mass function}
\author[B.K. Gibson \& F. Matteucci]{B.K. Gibson$^1$ \&
F. Matteucci$^2$ \\
$^1$ Mount Stromlo \& Siding Spring Observatories, Australian National 
University, Weston Creek P.O., Weston, ACT, Australia  2611 \\
$^2$ Scuola Internazionale Superiore di Studi Avanzati, Via Beirut 2-4,
Trieste, Italy  34013}
\maketitle
\begin{abstract}
Chemical and photometric models of elliptical galaxies with infall of
primordial gas, and subsequent ejection of processed material via galactic
winds, are described.  Ensuring that these models are consistent with
the present-day colour-luminosity relation \it and \rm the measured
intracluster medium (ICM) abundances, we demonstrate that the
initial mass function (IMF) must be significantly flatter 
(\ie $x\approx 0.80$) than the canonical Salpeter slope (\ie $x\approx 1.35$).  
Such a ``top-heavy'' IMF is in agreement with the earlier conclusions based
upon closed-box models for elliptical galaxy evolution.  On the other hand, the
top-heavy IMF, in conjunction with these semi-analytic infall models,
predicts an ICM gas mass which exceeds that observed by
up to a factor three, in contrast with the
canonical closed-box models.  Time and position-dependent IMF formalisms 
\it may \rm
prove to be a fruitful avenue for future research, but those presently
available in the literature are shown to be inconsistent with several
important observational constraints.
\end{abstract}



\section{Introduction}
\label{introduction}

Supernovae (SNe)-driven winds, and in particular their role in
setting the timescale for the cessation of bulk star formation, have long been
recognised as important components of elliptical galaxy formation/evolution
models.  Such wind models provide a natural framework in which the intrinsic
elliptical galaxy
colour-metallicity-luminosity (CML) relationships can be established
(\eg Larson 1974; Arimoto \& Yoshii 1987; Gibson 1996a,b).  An inescapable
consequence of such SNe-driven winds is the enrichment of the intergalactic
medium with the products of SNe nucleosynthesis (\eg Larson \& Dinerstein 1975).

Such wind ``pollution'' is particularly evident in the hot intracluster medium
(ICM) of elliptical-rich galaxy clusters.  Early, Type II SNe-dominated
winds are the favoured mechanism at play here, as evidenced by the strong
$\alpha$-element enhancement signatures in this hot gas (\eg [O/Fe]$\simgt
+0.2$ -- Mushotzky \etal 1996).  The absolute mass of iron in a cluster's ICM
can be shown to be related to the host cluster's integrated V-band luminosity
tied up in ellipticals; \ie
\begin{equation}
M_{\rm Fe}^{\rm ICM}\approx 0.02 L_{\rm V}^{\rm E},
\label{eq:eqn1}
\end{equation}
\noindent
where the mass and luminosity are in solar units (Arnaud 1994).
Similarly, a cluster's ICM gas mass $M_{\rm g}^{\rm ICM}
$ can be related to its early-type galaxy V-band luminosity by
\begin{equation}
M_{\rm g}^{\rm ICM}\approx 20\rightarrow 50  L_{\rm V}^{\rm E},
\label{eq:eqn2}
\end{equation}
\noindent
with loose groups and rich clusters populating the low and high luminosity
extrema, respectively.
While important,\footnote{For example, because of the $\alpha$-element
overabundance signature, late-time Type Ia SNe-driven wind 
models are no longer a viable \it dominant \rm ICM-pollution mechanism.} 
these ICM constraints
do not, by themselves, provide much information on the underlying star
formation or initial mass function (IMF) formalisms.

Previous papers in this series (Matteucci \& Gibson 1995; Gibson \& Matteucci
1997) have adopted the classic \it closed-box model \rm
for a galaxy's evolution and
concluded that IMFs significantly flatter-than-Salpeter's (1955) canonical
power-law slope $x=1.35$ are required, in order to fit the ICM constraints
noted above.  Similar conclusions have been
drawn by David, Forman \& Jones (1991)
Zepf \& Silk (1996) and Loewenstein \& Mushotzky (1996).

In the past year though, interest has been piqued by the adoption of gas 
\it infall \rm models for ellipticals (\eg Tantalo \etal 1996; Kodama \&
Arimoto 1997), in lieu of the aforementioned closed-box
scenarios.  Such studies retain the SNe-driven wind framework but
consciously restrict themselves
to the spectrophotometric properties of their model galaxies,
neglecting the consequences for enriching the ICM.  Because there exists a level
of degeneracy in the present-day CML relationships, due to uncertainties in the
IMF (\eg Table 8 of Gibson 1997), star formation methodology, 
SNe energetics, etc, 
Tantalo \etal and Kodama \& Arimoto used, partially for convenience,
IMF power law slopes close to the Salpeter (1955) value -- the
former, $x=1.35$; the latter, $x=1.20$.

The time is now ripe to assess the infall model in light of the heretofore
neglected ICM constraints -- in particular, the abundance ratios and absolute
gas masses -- in order to better appreciate its strengths and weaknesses.
What follows represents the first 
application of gas infall models to the ICM constraints, with answering the
question -- \it in the presence of gas infall, is the IMF in cluster
ellipticals dominated by high-mass stars (\ie top-heavy), or does it follow the
more canonical Salpeter distribution? \rm -- being the highest priority.

To this end,
in Section \ref{compare} we remind the reader of the primary motivation for
considering the infall model in the first place.
Following this, in Section \ref{analysis}, we
present a grid of models which parallels that
of Kodama \& Arimoto (1997).  We follow their numerical 
formalism, for convenience sake,
although its similarity to that of Tantalo \etal's (1996) means our
conclusions are independent of this choice.  
As we will show, the
favoured Kodama \& Arimoto grid, with IMF slope $x\approx 1.2$,
underproduces ICM
iron by approximately a factor of five.  We next present an alternate grid of
infall models consistent with the ICM abundance constraints.  As discussed in
Section \ref{analysis}, consistent with the results of earlier closed-box
models, only IMF slopes significantly flatter-then-Salpeter (\ie
$x\simlt 0.8$) were capable of recovering the galaxy CML relations \it and \rm
the ICM abundance constraints, albeit at the expense of the ICM gas mass
constraint.
Our results are summarised in Section \ref{summary}.

\section{Why infall models?}
\label{compare}

Recall that our earlier papers (Matteucci \& Gibson 1995; Gibson \& Matteucci
1997) adopted the straightforward closed-box model for galaxy evolution.
In particular, the star formation rate $\psi$ was assumed to vary directly with
the gas mass, as
\begin{equation}
\psi(t) = \nu M_{\rm g}(t),
\label{eq:eqn3}
\end{equation}
\noindent
with the timescale for star formation $\nu$ varying in such a way that the
present-day CML relations were properly established.  In the closed-box model,
star formation is a maximum at $t=0$, steadily decreasing thereafter.
Gibson \& Matteucci
(1997) showed that one can successfully recover the [(V-K),L$_{\rm V}$] relation
for ellipticals, and still honour all of the ICM abundance constraints, within
the closed-box model framework, \it but \rm only for IMF slopes significantly
flatter-than-Salpeter's (1955) -- \eg a slope $x\approx 1$, as opposed to
Salpeter's canonical $x=1.35$, was necessary.

It has since become apparent (\eg Tantalo \etal 1996; Kodama \& Arimoto 1997)
though that this same closed-box model for
ellipticals does not appear to 
have the same success in recovering the
\it ultraviolet (UV)-optical \rm
CML relations, a point not explored in the Gibson
\& Matteucci (1997) study.  The reason for this failure
is the overproduction of low-metallicity (\ie Z=0)
stars during the initial intense star
formation regime; the signature of such a component is not seen in the
integrated spectra of old stellar populations (Worthey, Dorman \& Jones 1996).

Several alternate star formation
formalisms, aimed specifically at avoiding this low-metallicity overproduction,
have since been published -- the previously noted
Tantalo \etal (1996) and Kodama \& Arimoto (1997) studies being the most
noteworthy.  These infall models are similar in
spirit to those designed to avoid the analogous G-dwarf problem 
encountered by solar neighbourhood-enrichment models (\eg Tinsley 1980),
although for ellipticals, the infall timescales are, of course, substantially
different to the slower accretion rates expected for spirals.  Following 
Tantalo \etal and Kodama \& Arimoto,
we adopt an infall rate which parallels the free-fall timescale.
Star formation, in both models, is assumed to follow equation 
\ref{eq:eqn3},
but instead, with an initial gas mass of zero, increasing in accordance with
the assumed gas infall rate law.  The accompanying
rapid increase in global metallicity means that very few low-metallicity stars
are formed, avoiding the G-dwarf problem.  

In summary, gas infall models for elliptical galaxies have been considered as
viable alternatives to the classic closed-box model, primarily because they
provide a simple solution to the overproduction of low metallicity stars which
plagues the closed-box models - such a G-dwarf problem manifests itself most
clearly in the closed-box model's problem in recovering the
UV-optical CML relations.

\section{Analysis}
\label{analysis}

\subsection{The models}
\label{models}

Using the photo-chemical evolution code of Gibson (1996a,b;1997), 
we constructed a
grid of seven elliptical galaxy models consistent with the ``metallicity
sequence'' of Kodama \& Arimoto (1997, hereafter KA97).  In particular, KA97
adopted an IMF slope mildly flatter (\ie $x=1.20$) than Salpeter's (1955)
$x=1.35$, with a mass range of $0.1\rightarrow 60.0$ M$_\odot$.  The cessation
of star formation (\ie at time $t_{\rm GW}$) was treated as a free parameter,
and varied to ensure recovery of the present-day elliptical galaxy CML
relations.\footnote{It should be noted that KA97's models are
systematically redder in (V-K) than ours, by $\sim 0.06$ mag.  Tracing the
source of this discrepancy is not possible at this time, as their simple
stellar population (SSP)
colours are not currently available.  Our grid is based upon the published
Bertelli \etal (1994) isochrones, with the low mass extensions outlined in
Gibson (1996a).  For the alternate grid of models to be
discussed shortly, we ensure self-consistency by maintaining this systematic
0.06 magnitude offset in (V-K).}
Reasonable values for the parameters governing the timescales for
star formation and gas infall were adopted (\ie 0.1 Gyr), independent of galaxy 
model.  The adopted nucleosynthetic yields are described in Gibson \&
Mould (1997).

The first block of models in Table 1 represent our pseudo-KA97 grid
(\ie those labelled $x=1.20$);
unlike the data presented in their Table 2, to which the reader is referred to
for additional parameters
not relevant to the discussion at hand, we have included
the masses of gas, oxygen, and iron ejected to the ICM, for each model (in
units of $10^9$ M$_\odot$, $10^6$ M$_\odot$, and $10^6$ M$_\odot$,
respectively).  The initial gas mass reservoir associated with each model (\ie
$M_{\rm T}$, in units of $10^9$ M$_\odot$) is also provided in column 2, as it
was not included in KA97's Table 2.  Column 7 shows the residual reservoir gas
mass at the time of galactic gas expulsion $t_{\rm GW}$, again in units of
$10^9$ M$_\odot$.

\begin{table*}[hpt]
\caption[]{\label{tbl:models}
Elliptical galaxy models constructed to conform to the \it metallicity
sequence \rm of Kodama \& Arimoto (1997) -- \ie the block containing models
with an IMF slope $x=1.20$.  The second block of models (\ie IMF slope
$x=0.80$) was constructed \it a posteriori \rm to ensure compatibility with
Arnaud's (1994) cluster V-band luminosity-ICM iron mass relationship (equation
1).  
The masses of gas, oxygen, and iron ejected at time
to the ICM (columns 4, 5, and 6) at time $t_{\rm GW}$
are in units of $10^9$ M$_\odot$, $10^6$ M$_\odot$, and $10^6$ M$_\odot$,
respectively.  Column 7 is the mass of gas remaining in the initial reservoir
at time $t_{\rm GW}$, also in units of $10^9$ M$_\odot$.  See text for details.}
\begin{center}
\begin{tabular}{rrrrrrr}
\vspace{2.0mm}
$M_{\rm V}$ & $M_{\rm T}$ & $[<{\rm Z}>]_{\rm V}$ & $M_{\rm g}(t_{\rm GW})$ &
$M_{\rm O}(t_{\rm GW})$ & $M_{\rm Fe}(t_{\rm GW})$ & $M_{\rm g}^{\rm
res}(t_{\rm GW})$ \\
\multicolumn{7}{c}{$x=1.20$} \\
-22.91 & 1000 & +0.16$\quad$ &  79$\quad$  &  2832$\quad$ &  218$\quad$ &   9$\quad$ \\
-21.99 &  500 & -0.06$\quad$ & 102$\quad$  &  2508$\quad$ &  168$\quad$ &   20$\quad$ \\
-21.05 &  250 & -0.19$\quad$ &  75$\quad$  &  1369$\quad$ &   87$\quad$ &   24$\quad$ \\
-19.84 &  100 & -0.32$\quad$ &  36$\quad$  &   516$\quad$ &   32$\quad$ &   16$\quad$ \\
-18.86 &   50 & -0.44$\quad$ &  20$\quad$  &   224$\quad$ &   13$\quad$ &   12$\quad$ \\
-17.90 &   25 & -0.53$\quad$ &  10$\quad$  &    93$\quad$ &    5$\quad$ &   8$\quad$ \\
-16.66 &   10 & -0.64$\quad$ &   4$\quad$  &    30$\quad$ &    2$\quad$ &   4$\quad$ \\
\multicolumn{7}{c}{$x=0.80$} \\
-22.44 & 3000 & -0.12$\quad$ & 1404$\quad$ & 34660$\quad$ & 1725$\quad$ & 904$\quad$ \\
-20.78 & 1000 & -0.31$\quad$ &  427$\quad$ &  7353$\quad$ &  336$\quad$ & 427$\quad$ \\
-20.32 &  700 & -0.34$\quad$ &  294$\quad$ &  4789$\quad$ &  216$\quad$ & 311$\quad$ \\
-19.08 &  300 & -0.48$\quad$ &  114$\quad$ &  1439$\quad$ &   61$\quad$ & 157$\quad$ \\
\end{tabular}
\end{center}
\end{table*}

\begin{figure}
\centering
\vspace{8.5cm}
\includegraphics{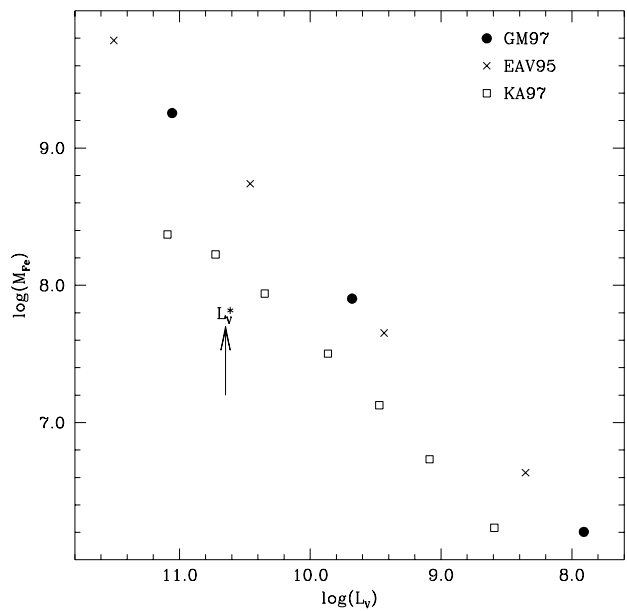}
\caption[]{Mass of iron ejected to the ICM as a function of
present-day V-band luminosity.  The models of Gibson \& Matteucci (1997, GM97),
Elbaz \etal (1995, EAV95), and Kodama \& Arimoto (1997, KA97) are shown.  The
characteristic cluster galaxy V-band luminosity L$_{\rm V}^\ast$ is
indicated.
\label{fig:fig1}}
\end{figure}

Of immediate concern is the mass of iron ejected from
each of the model galaxies (column 6 of Table 1).  In Figure \ref{fig:fig1} we
show this mass of iron M$_{\rm Fe}$, for the Kodama \& Arimoto (1997) models of
Table 1, as a function of the present-day V-band luminosity (open squares).
For comparison, the predictions of Gibson \& Matteucci (1997, their ``standard
model'') and Elbaz, Arnaud \& Vangioni-Flam (1995) are also shown, with filled
circles and crosses, respectively.  We stress that the latter two models have
been specifically designed to honour Arnaud's (1994) cluster V-band
luminosity--ICM iron mass relation (equation \ref{eq:eqn1}).\footnote{While
it is true that both Gibson \& Matteucci's (1997) and Elbaz \etal's (1995)
models obey equation \ref{eq:eqn1}, only the former is consistent with the
optical-infrared elliptical galaxy CML relations (Gibson 1996a).}
It should be readily apparent from Figure \ref{fig:fig1} that Kodama
\& Arimoto's models systematically underproduce (by a factor of $\sim 5$)
the necessary iron to recover Arnaud's relationship (equation
\ref{eq:eqn1}).  

\subsection{Infall and the IMF}
\label{imf}

Our next step was to investigate mechanisms whereby Arnaud's (1994)
relationship \it could \rm
be honoured, but from within the infall framework.  Varying
the timescale for star formation, the gas infall rate, or the yield selection, 
does not alter the conclusion of the previous paragraph -- \ie the models still
underproduce iron.\footnote{Invoking a later-time Type Ia SNe-driven wind
component of such magnitude as to fully
recover the ICM iron mass-cluster V-band
luminosity relationship (equation \ref{eq:eqn1}), drives the predicted ICM
[O/Fe] to $\sim -0.5$, a factor of $\sim 5\rightarrow 10\times$ lower than
observed (Mushotzky \etal 1996).}
\it The only ingredient which could be varied, and recover the $M_{\rm Fe}^{\rm
ICM}$--$L_{\rm V}^{\rm E}$
relationship successfully, was the IMF
slope.\rm\footnote{Bimodal IMFs are also a possibility, but in their simplest
form (Elbaz \etal 1995) do not recover the observed CML relations (Gibson
1996a).}

\begin{figure}
\centering
\vspace{8.5cm}
\includegraphics{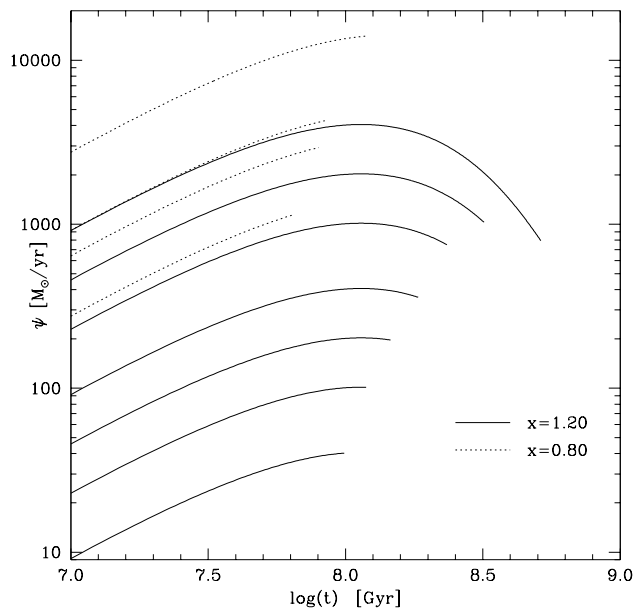}
\caption[]{Evolution of the star formation rate $\psi$ as a function of
time $t$, for the seven ``metallicity-sequence'' models of Kodama \& Arimoto
(1997), with IMF slope $x=1.20$ (solid curve).  Also shown are the parallel grid
of models with slope $x=0.80$ (dotted curve), constructed \it a posteriori \rm
to recover Arnaud's (1994) cluster luminosity-ICM iron mass relation.  
The galactic mass of the
respective models decreases from top-to-bottom, as per the corresponding
entries in Table 1.
\label{fig:fig2}}
\end{figure}

Keeping all other input parameters the same as for the $x=1.20$ grid of
models, we varied the IMF slope $x$ and the initial gas reservoir mass $M_{\rm
T}$, until a grid was generated which honoured Arnaud's (1994) $M_{\rm Fe}^{\rm
ICM}-L_{\rm V}^{\rm E}$ relationship (equation \ref{eq:eqn1})
\it and \rm retained consistency with
the elliptical galaxy [(V-K),L$_{\rm V}$] relation.  
An IMF slope $x=0.80$ was found to
be necessary, in order to recover the mean of this relationship.  
A subset of these models are shown in the second block of Table 1.  
The galactic wind epochs for these models ranged from 120 to 65 Myrs (from most
to least luminous).
For comparison,
Figure \ref{fig:fig2} shows how the 
star formation rate evolves for this subset (dotted curves), as well as those
for the standard $x=1.20$ models (solid curves) of Table 1.

\subsection{Potential problems with the infall scenario}
\label{problems}

Besides our conclusion regarding the necessarily steeper-than-Salpeter (1955)
IMF slope of $x\approx 0.80$, there are two other points which should be drawn
from the model grid of Table 1.\footnote{The mass-to-blue light ratios
(M/L$_{\rm B}$) for the $x=0.80$ models are $\sim 10\rightarrow 15$\% lower than
the corresponding $x=1.20$ grid.  While interesting, the uncertainties in
deriving the observational M/L$_{\rm B}$--L$_{\rm B}$
relation make any further discussion regarding this difference moot.}
First, the standard $x=1.20$ grid \it already
\rm lies consistently
$\sim 0.1\rightarrow 0.3$ dex below the mean observed $\bigl[[<Z>]_{\rm
V}$,M$_{\rm V}\bigr]$ relation (\eg see Figure 1 of Gibson \& Matteucci 1997).
The new $x=0.80$ grid, which is now consistent with the ICM constraint of
equation \ref{eq:eqn1}, only exacerbates this shortcoming, being a \it further
\rm $\sim 0.1$ dex removed from the mean.  Second, and perhaps more
importantly, the flatter IMF grid
requires an initial gas mass reservoir $\sim 5\times$ more massive than that
for the steeper IMF grid, for galactic models of the same present-day
luminosity.

Following Section 7 of Elbaz \etal (1995), we then use the models of Table 1
to find that the \it predicted \rm ICM gas mass to cluster luminosity is
\begin{equation}
M_{\rm T}^{\rm ICM}\approx 8 L_{\rm V}^{\rm E},
\label{eq:eqn4}
\end{equation}
\noindent
for the $x=1.20$ grid, and
\begin{equation}
M_{\rm T}^{\rm ICM}\approx 55 L_{\rm V}^{\rm E},
\label{eq:eqn5}
\end{equation}
\noindent
for the $x=0.80$ grid.  This predicted ICM gas mass includes the mass of gas
ejected at $t_{\rm GW}$ (column 4 of Table 1) and the residual, unincorporated,
reservoir gas (column 7).  An additional component from late-time (\ie
$t> t_{\rm GW}$) winds or ram pressure stripping can be included but does not
alter the coefficients shown in equations \ref{eq:eqn4} and \ref{eq:eqn5}.
The coefficients can be reduced by a factor of $\sim 2$ to recover the relation
based upon the ejected gas mass at $t_{\rm GW}$ alone.

For the $x=1.20$ IMF, equation \ref{eq:eqn4}
implies that $\sim 40\rightarrow 15$\% (loose groups to rich clusters)
of the total ICM gas (\ie equation \ref{eq:eqn2}) has
been accounted for; $\sim 60$\% of this has been processed and ejected from
cluster ellipticals, and $\sim 40$\% is associated with residual reservoir-gas.
\footnote{These percentages have associated
errors of $\sim 20$\%, as we have fit simple
power-laws to the gas mass--cluster luminosity relations in order to follow the
procedure laid out in Section 7 of Elbaz \etal (1995).  Strictly, a numerical
integration should be performed, as discussed already in Gibson \& Matteucci
(1997), but for simplicity's sake we adhere to the Elbaz \etal formalism.}
The remaining $\sim 60\rightarrow 85$\% (loose groups to 
rich clusters) of the
observed ICM gas must therefore be assigned to some gas component which simply
does not partake in the star formation process (either ``actively'' through
galactic wind ejection, or ``passively'', by being associated with a specific
galactic halo's initial gas reservoir).
This is perfectly in keeping with the results of
Matteucci \& Vettolani (1988), David \etal (1991),
Elbaz \etal (1995), and Gibson \& Matteucci
(1997).  

On the other hand, equation \ref{eq:eqn5}, coupled with
the $x=0.80$ grid, implies that $\sim 300\rightarrow
100$\% (again, loose groups to rich clusters) of the total ICM gas (\ie
equation \ref{eq:eqn2}) has been accounted for!  Again, this gas can be
considered to be $\sim 50$\% ``processed-and-ejected'' and $\sim 50$\%
``initial gas reservoir-associated'' -- no other cluster gas constituent need
be invoked.  The problem lies, of course, in that for any system ``poorer''
than the richest clusters, there is an accompanying overproduction of gas --
\ie \it the grid of infall models constructed with the most conservative input
parameters require, and yield, more gas than is actually observed, for all but
the richest clusters.  \rm Some of this 
discrepancy may be alleviated if the galactic
winds are successful in overcoming not only an individual galaxy's potential
well, but that of the group's as well, dispersing some fraction of the
``excess'' gas to the general intergalactic medium.

\subsection{Variable IMFs?}
\label{variable}

Because of the closed-box model's failure to properly recover the UV-optical
CML relations (recall
Section \ref{compare}) and the infall model's problems with gas
``overproduction'', it might be tempting, by process of elimination, to
support the notion of time (and 
position)-variable IMFs, a l\`a 
Padoan, Nordlund \& Jones (1997).  Detailed models based upon Padoan \etal's
``universal'' IMF formalism will be presented elsewhere, but we would be remiss
if we did not at least temper such ``default'' support, by noting two of the
weaknesses in the (currently) best available variable-IMF 
elliptical galaxy models.

Chiosi \& Bressan (1997) have presented their first (preliminary) grid of 
elliptical galaxy models, based upon the Padoan \etal (1997) variable-IMF
formalism.  These models 
were designed primarily to satisfy one projection of the
fundamental plane (specifically, M/L$_{\rm B}$ versus M$_{\rm B}$), 
while still retaining consistency with the underlying galaxy
CML relations and observed $\alpha$-element \it stellar population \rm
overabundance.  While successful in satisfying these constraints, there are two
equally important observational constraints which are clearly violated.

First, local dwarf ellipticals such as Fornax, NGC 185, and NGC
205, each host populations of planetary nebulae with [O/Fe] typically $\sim
+0.2$, and ages presumably on 
the order of several Gyrs (Richer, McCall \& Arimoto 1997).   An inescapable
result of the Chiosi \& Bressan (1997) ``canonical''
dwarf elliptical (dE)
model (see their Figure 2) is that dE planetary nebulae, of this typical age,
should have [O/Fe] of $\simlt -1.4$, a factor of $\sim 40$ lower than that
observed.

Second, and perhaps more important, is the predicted ICM [O/Fe] from the Chiosi
\& Bressan (1997) grid.  As their dwarf ellipticals never develop galactic
winds, it is really only the $\sim$L$_\ast$ galaxies which can 
contribute to the observed [O/Fe]$\simgt +0.2$ (Mushotzky \etal 1996).  Again,
referring to their Figure 1, we can see that their $\sim$L$_\ast$ ellipticals
develop galactic winds after $\sim 3.5$ Gyrs, by which point the ISM [O/Fe] has
been reduced to $\sim -1$ (their Figure 2).  \it 
This factor of $\sim 10\rightarrow 20$ discrepancy
between observed and theoretical ICM [O/Fe] points to a significant
deficiency in the existing elliptical galaxy variable-IMF models. \rm

Third, the Chiosi \& Bressan (1997) $\sim$L$_\ast$ model (again, coupling
their Figure 1 with the adopted star formation formalism) 
shows a star formation
rate, in the core alone (where $\simlt 10$\% of the galaxy's luminous mass
lies), ranging from 
$\psi\sim 6000\rightarrow 1500$ M$_\odot$/yr from redshifts of
$\sim 5$ to $\sim 1$.  The total \it global \rm star formation rate is $\sim
2\rightarrow 3\times$ this \it core \rm value.  Such star formation rates at
$z\sim 1$ (\ie $\psi\simgt 3000$ M$_\odot$), regardless of IMF,
are clearly at odds with the
cosmological number counts (\eg Charlot \& Silk 1995), although a more formal
analysis, in a cosmological context, is postponed for the time being.

To end on a positive note though,
the above arguments do not necessarily preclude the
possibility that future, more sophisticated, variable-IMF
models, may reconcile \it all \rm of the galaxy CML relations \it and
\rm ICM constraints.  It should be re-iterated that the Chiosi \&
Bressan (1997) results are still preliminary, and that the full parameter space
still needs to be explored.  What is clear though, is that the models presented
\it thus far \rm are not necessarily the panacea they might at
first appear to be.

\section{Summary}
\label{summary}

We have constructed a grid of gas infall models of elliptical galaxies, based
upon the favoured sequence of Kodama \& Arimoto (1997).  For an IMF power-law
slope $x=1.20$, the models are entirely consistent with the observed,
present-day, colour-luminosity relations for cluster ellipticals.  While
previous closed-box models were equally successful in recovering the
optical-infrared relations, the infall models do have the advantage of a better
fit to the ultraviolet-optical relations (albeit, at the expense of a poorer
fit to the metallicity-luminosity relation).  We assess qualitatively
the ability of Padoan \etal's (1997) variable-IMF scenario, as adopted by Chiosi
\& Bressan (1997), to circumvent the closed-box/infall ``problems'', and
conclude that the extent models admittedly minimise some of these problems, \it
but \rm only at the expense of introducing new ones.

For the first time, we have employed infall models as input
to simple cluster enrichment scenarios.  For the published grids, which appear
to favour IMF slopes similar to that of the canonical Salpeter (1955) value
(\ie $x\approx 1.2\rightarrow 1.4$), an inescapable conclusion is that the
predicted mass of iron ejected to a cluster's ICM is a factor of $\sim 5$ below
that observed.  Only by substantially ``flattening'' the IMF (to $x\approx
0.8$), were we able to construct an ``infall'' grid which was consistent with
\it both \rm the galactic colour-luminosity relations and ICM iron mass-cluster
luminosity relation, albeit at (a) the expense of modestly worsening the 
galactic metallicity-luminosity relation, and (b) overproducing ICM gas for
all but the richest clusters.  
This latter point may be indicative of a
fundamental flaw in the infall picture, but we prefer to reserve judgment
on this point until more sophisticated models have been explored.

In the meantime though, we are able to conclude that,
by adopting an already accepted (\eg Tantalo \etal 1996; Kodama \& Arimoto
1997) formalism for elliptical galaxy evolution based upon gas infall, 
the IMF must necessarily favour the formation of massive stars
(\ie be ``top-heavy'') 
if the models are to conform to the present-day CML relations,
as well as the ICM abundance constraints.  This agrees with our earlier studies
based upon the more conventional closed-box
model (Matteucci \& Gibson 1995; Gibson \& Matteucci 1997).  It is
encouraging that two such disparate approaches (\ie closed-box versus infall)
to modeling the evolution of
ellipticals \it both \rm lead to the same conclusion
regarding the veracity of the top-heavy IMF hypothesis.

\section*{ACKNOWLEDGEMENTS}
We are grateful to Tadayuki Kodama and David Woods,
for a number of useful correspondences.
BKG acknowledges the financial support of an NSERC Postdoctoral Fellowship.




\begin{thebibliography}{}

\bibitem[Arimoto \& Yoshii 1987]{AY87}
Arimoto, N. \& Yoshii, Y. 1987,
\aap, 173, 23

\bibitem[Arnaud 1994]{A94}
Arnaud, M. 1994,
Clusters of Galaxies, ed. F. Durret, A. Mazure \& J. Tran Thanh Van,
Gif-sur-Yvette: Editions Frontieres, 211

\bibitem[Bertelli \etal 1994]{BBCFN94}
Bertelli, G., Bressan, A., Chiosi, C., Fagotto, F. \& Nasi, E. 1994,
A\&AS, 106, 275

\bibitem[Charlot \& Silk 1995]{CS95}
Charlot, S. \& Silk, J. 1995,
\apj, 445, 124

\bibitem[Chiosi \& Bressan 1997]{CB97}
Chiosi, C. \& Bressan, A. 1997,
Galaxy Scaling Relations: Origins, Evolution and Applications, ed. da Costa, L.
\etal, in press

\bibitem[David \etal 1991]{DFJ91}
David, L.P., Forman, W. \& Jones, C. 1991,
\apj, 380, 39

\bibitem[Elbaz \etal 1995]{EAV95}
Elbaz, D., Arnaud, M. \& Vangioni-Flam, E. 1995,
\aap, 303, 345

\bibitem[Gibson 1996a]{G95b}
Gibson, B.K. 1996a,
\mnras, 278, 829

\bibitem[Gibson 1996b]{G96}
Gibson, B.K. 1996b,
\apj, 468, 167

\bibitem[Gibson 1997]{G97}
Gibson, B.K. 1997,
\mnras, in press

\bibitem[Gibson \& Matteucci 1997]{GM96}
Gibson, B.K. \& Matteucci, F. 1997,
\apj, 475, 47

\bibitem[Gibson \& Mould 1997]{GM97}
Gibson, B.K. \& Mould, J.R. 1997,
\apj, 482, 98

\bibitem[Kodama \& Arimoto 1997]{KA97}
Kodama, T. \& Arimoto, N. 1997,
\aap, 320, 41

\bibitem[Larson 1974]{L74}
Larson, R.B. 1974,
\mnras, 169, 229

\bibitem[Larson \& Dinerstein 1975]{LD75}
Larson, R.B. \& Dinerstein, H.L. 1975,
\pasp, 87, 911

\bibitem[Loewenstein \& Mushotzky 1996]{LM96}
Loewenstein, M. \& Mushotzky, R.F. 1996,
\apj, 466, 695

\bibitem[Matteucci \& Gibson 1995]{MG95}
Matteucci, F. \& Gibson, B.K. 1995,
\aap, 304, 11

\bibitem[Matteucci \& Vettolani 1988]{MV88}
Matteucci, F. \& Vettolani, G. 1988,
\aap, 202, 21

\bibitem[Mushotzky \etal 1996]{Mush96}
Mushotzky, R., Loewenstein, M., Arnaud, K.A., Tamura, T., Fukazawa, Y.,
Matsushita, K., Kikuchi, K. \& Hatsukade, I. 1996,
\apj, 466, 686

\bibitem[Padoan, Nordlund \& Jones 1997]{PNJ97}
Padoan, P., Nordlund, A.P. \& Jones, B.J.T. 1997,
\mnras, 288, 145

\bibitem[Salpeter 1955]{S55}
Salpeter, E.E. 1955,
\apj, 121, 161

\bibitem[Tantalo \etal 1996]{TCBF96}
Tantalo, R., Chiosi, C., Bressan, A. \& Fagotto, F. 1996,
\aap, 311, 361

\bibitem[Tinsley 1980]{T80}
Tinsley, B.M. 1980,
Fund. Cosm. Phys., 5, 287

\bibitem[Worthey, Dorman \& Jones 1996]{WDJ96}
Worthey, G., Dorman, B \& Jones, L.A. 1996,
\aj, 112, 948

\bibitem[Zepf \& Silk 1996]{ZS96}
Zepf, S.E. \& Silk, J. 1996,
\apj, 466, 114

\end{thebibliography}
\end{document}